\documentclass[12pt,preprint]{aastex}
\newcommand{\kms}{\mbox{km s$^{-1}$}}

\newcommand{\Msun}{\mbox{$M_{\sun}$}}
\newcommand{\Lsun}{\mbox{$L_{\sun}$}}
\newcommand{\skipthis}[1]{}
\newcommand{\hii}{\mbox{\ion{H}{2}}}

\shortauthors{Garay et al.}
\shorttitle{Dust in massive star forming regions}
\slugcomment{submitted to {\em The Astrophysical Journal}}
  
\begin{document}

%Draft \today

%%%%%%%%%%%%%%%%%%%%%%%%%%%%%%%%%%%%%%%%%%%%%%%%%%%%%%%%%%%%%%%%%%%%%%%%%
% TITLE and AUTHORS
%%%%%%%%%%%%%%%%%%%%%%%%%%%%%%%%%%%%%%%%%%%%%%%%%%%%%%%%%%%%%%%%%%%%%%%%%

\title{A multiwavelength study of young massive star forming regions: 
II. The dust environment}
  
\author{Guido Garay}
\affil{Departamento de Astronom\'{\i}a, Universidad de Chile,
Casilla 36-D, Santiago, Chile}
% guido@das.uchile.cl

\author{Diego Mardones}
\affil{Departamento de Astronom\'{\i}a, Universidad de Chile,
Casilla 36-D, Santiago, Chile}
% mardones@das.uchile.cl

\author{Kate J. Brooks}
\affil{Australia Telescope National Facility, P.O. Box 76, Epping NSW 1710,
  Australia}

\author{Liza Videla}
\affil{Departamento de Astronom\'{\i}a, Universidad de Chile,
Casilla 36-D, Santiago, Chile}

\and

\author{Yanett Contreras}
\affil{Departamento de Astronom\'{\i}a, Universidad de Chile,
Casilla 36-D, Santiago, Chile}

%%%%%%%%%%%%%%%%%%%%%%%%%%%%%%%%%%%%%%%%%%%%%%%%%%%%%%%%%%%%%%%%%%%%%%%%%
% Abstract
%%%%%%%%%%%%%%%%%%%%%%%%%%%%%%%%%%%%%%%%%%%%%%%%%%%%%%%%%%%%%%%%%%%%%%%%%
\begin{abstract}

We present observations of 1.2-mm dust continuum emission, made with the 
Swedish ESO Submillimeter Telescope, towards eighteen 
luminous IRAS point sources, all with colors typical of compact \hii\ regions 
and associated with CS(2$\rightarrow$1) emission, thought to be representative 
of young massive star forming regions. Emission was detected toward all the 
IRAS objects. We find that the 1.2-mm sources associated with them have 
distinct physical parameters, namely sizes of $\sim0.4$ pc, dust temperatures 
of $\sim30$ K, masses of $\sim2\times10^3$ M$_\odot$, column densities of 
$\sim3\times10^{23}$ cm$^{-2}$, and densities of $\sim4\times10^{5}$ cm$^{-3}$. 
We refer to these dust structures as massive and dense cores. Most of the 
1.2-mm sources show single-peaked structures, several of which exhibit 
a bright compact peak surrounded by a weaker extended envelope. The observed 
radial intensity profiles of sources with this type of morphology 
are well fitted with power-law intensity profiles with power-law indices 
in the range $1.0-1.7$. This result indicates that massive and dense cores
are centrally condensed, having radial density profiles with power-law indices 
in the range $1.5-2.2$. We also find that the UC HII regions detected with ATCA 
towards the IRAS sources investigated here (Paper I) are usually projected at 
the peak position of the 1.2-mm dust continuum emission, suggesting that 
massive stars are formed at the center of the centrally condensed massive 
and dense cores.

\end{abstract}


\vfill\eject

\section{Introduction}

The birth process and early evolution of massive stars is currently one of 
the most debated subjects in astrophysics. To achieve a better understanding of 
the formation processes of high-mass stars it is essential to know the physical 
conditions of the parental cloud which are thought to play a critical role in 
determining the formation mechanism.  Accordingly, during the last decade 
intense observational 
efforts have been carried out to determine the characteristics of the molecular 
gas (e.g., Plume et al. 1992, 1997; Juvela 1996; Shirley et al. 2003)
and dust (e.g., Beuther et al. 2002; Mueller et al. 2002; Fa\'undez et al. 2004; 
Williams et al. 2004) associated with massive star forming regions. 

In this regard we are carrying out a multi-wavelength study of a sample of 
18 luminous IRAS sources in the southern hemisphere thought to be representative 
of young massive star forming regions. The goal is to understand the physical 
and chemical differences between different stages of early evolution. The objects 
were taken 
from the Galaxy-wide survey of CS(2$\rightarrow$1) emission towards IRAS sources 
with IR colors typical of compact \hii\ regions (Bronfman, Nyman, \& May 1996). 
We selected sources based primarily on the observed CS(2$\rightarrow$1) line 
profiles; looking for self-absorbed lines consistent with inward or outward 
motions (e.g., Mardones 1998), and/or with extended line wings, possibly 
indicating the presence of bipolar outflows. In addition, the sources were
required to have IRAS 100$\mu$m fluxes greater than $10^3$ Jy and to be in the
southern hemisphere ($\delta < -20\arcdeg$). The luminosity of the IRAS sources,
computed using the IRAS energy distribution and distances derived by Bronfman
(2006, private communication) are in the range $1\times10^4 - 4\times10^5$ \Lsun, 
implying that they contain at least an embedded massive star.

Here we report the results of 1.2-mm continuum observations, made with the 
Swedish ESO Submillimeter Telescope with angular resolution of 24\arcsec, 
toward all objects in our sample. The 1.2-mm emission was mapped within regions 
of $\sim15\arcmin\times10\arcmin$ centered on the IRAS sources. The primary goal 
of these observations was to determine the characteristics and physical 
conditions of the dust and gas clouds harboring recently formed massive stars. 
The characteristics of the ionized gas associated with the IRAS 
sources in our sample is reported in the first paper of the series
(Garay et al. 2006; Paper I).

%%%%%%%%%%%%%%%%%%%%%%%%%%%%%%%%%%%%%%%%%%%%%%%%%%%%%%%%%%%%%%%%%%%%%%%%%
% OBSERVATIONS
%%%%%%%%%%%%%%%%%%%%%%%%%%%%%%%%%%%%%%%%%%%%%%%%%%%%%%%%%%%%%%%%%%%%%%%%%

\section{Observations}

The observations were made using the 15 m Swedish--ESO Submillimetre Telescope
(SEST) located on La Silla, Chile. The 1.2-mm continuum observations were made
using the 37-channel SEST Imaging Bolometer Array (SIMBA) during October, 2001 
and July 2002. The passband of the bolometers has an equivalent width
of $90$ GHz and is centered at $250$ GHz. The HPBW of a single element is
24\arcsec\ and the separation between elements on the sky is 44\arcsec. We
observed in the fast mapping mode, using a scan speed of 80\arcsec\ s$^{-1}$.
Each observing block consisted of 50 scan lines in azimuth of length 800\arcsec\
and separated in elevation by 8\arcsec, giving a map size of 400\arcsec\
in elevation. This block required $\sim$ 15 minutes of observing time. We 
observed $2-4$ blocks per source, achieving rms noise levels of typically $50$ 
mJy beam$^{-1}$. The data were reduced according to a standard procedure 
using the software package MOPSI, which included baseline subtraction and 
rejection of correlated sky-noise. Flux calibration was performed using a 
sky-opacity correction and a counts-to-flux conversion factor derived from maps 
of Uranus. Uncertainties in the absolute calibration and pointing accuracy are 
estimated at $20\%$ and 3\arcsec, respectively.
The observed sources are listed in Table~\ref{tbl-obssources}. The noise
level achieved in the images is given in column $4$.

%%%%%%%%%%%%%%%%%%%%%%%%%%%%%%%%%%%%%%%%%%%%%%%%%%%%%%%%%%%%%%%%%%%%%%%%%
% RESULTS
%%%%%%%%%%%%%%%%%%%%%%%%%%%%%%%%%%%%%%%%%%%%%%%%%%%%%%%%%%%%%%%%%%%%%%%%%

\section{Results}

Figures~\ref{fig-simba1}-\ref{fig-simba3} present maps of the 1.2-mm emission,
towards all the observed targets, within regions of typically $5'-7'$ in size. 
The filled stars and open triangles indicate, respectively, the peak position 
of the compact and more diffuse regions of ionized gas detected with ATCA 
within the mapped regions (Paper I). In a few panels we have also indicated, 
with open asterisks, the peak position of the sources detected in the {\sl Midcourse 
Space Experiment} (MSX; Price 1995) survey of the Galactic plane (Egan et al. 
1998) at 21.3 $\mu$m. We will refer to these as MSX-E band sources, 
most of which are likely to signpost the presence of embedded energy sources.

Continuum 1.2-mm emission was detected toward the eighteen IRAS sources. 
The observed parameters of the 1.2-mm sources as a whole are given in 
Table~\ref{tbl-obsparwhole}. Cols.(2) and (3) give their peak position. Cols.(4) 
and (5) give, respectively, the peak flux density and the total flux density, 
the later measured directly from the maps using the AIPS tasks IMEAN.
Col.(6) gives the deconvolved major and minor FWHM angular sizes determined 
by fitting a single Gaussian profile to the entire source.
Col.(7) gives a description of the source morphology. The {\it c+e} morphology 
is used to describe a source with a brightness distribution showing a bright 
compact peak surrounded by a weaker extended envelope. The criteria for 
classifying a source in this category is as follows. A single Gaussian fit 
has to produce distinctive residuals, consisting of a central positive peak 
surrounded by a ring of negative values and further out a ring of positive 
values. This is what is expected when a power law profile is fitted with a 
single Gaussian profile. On the other hand, two Gaussian fits -- a compact 
bright component plus an extended weaker component surrounding the former -- 
has to produce reasonable fits with the central position of both components 
being coincident within the observational errors. Thus the {\it c+e} sources 
should also be understood as sources in which the brightness distribution 
follows a power-law dependence with radius. For sources with multiple or 
complex morphology we fitted the brightness distribution using the minimum 
number of Gaussian components that produce residuals that are consistent 
with the noise level in the maps. The observed parameters of the fitted 
components are given in Table~\ref{tbl-obsparind}. Here flux densities and 
angular sizes corresponds to the values obtained by the fitting process using 
the AIPS task JMFIT. In what follows we discuss the characteristics of the 
1.2-mm emission toward each of the IRAS sources, individually. 

\subsection{Individual sources}

\begin{figure}
\epsscale{1.0}
\plotone{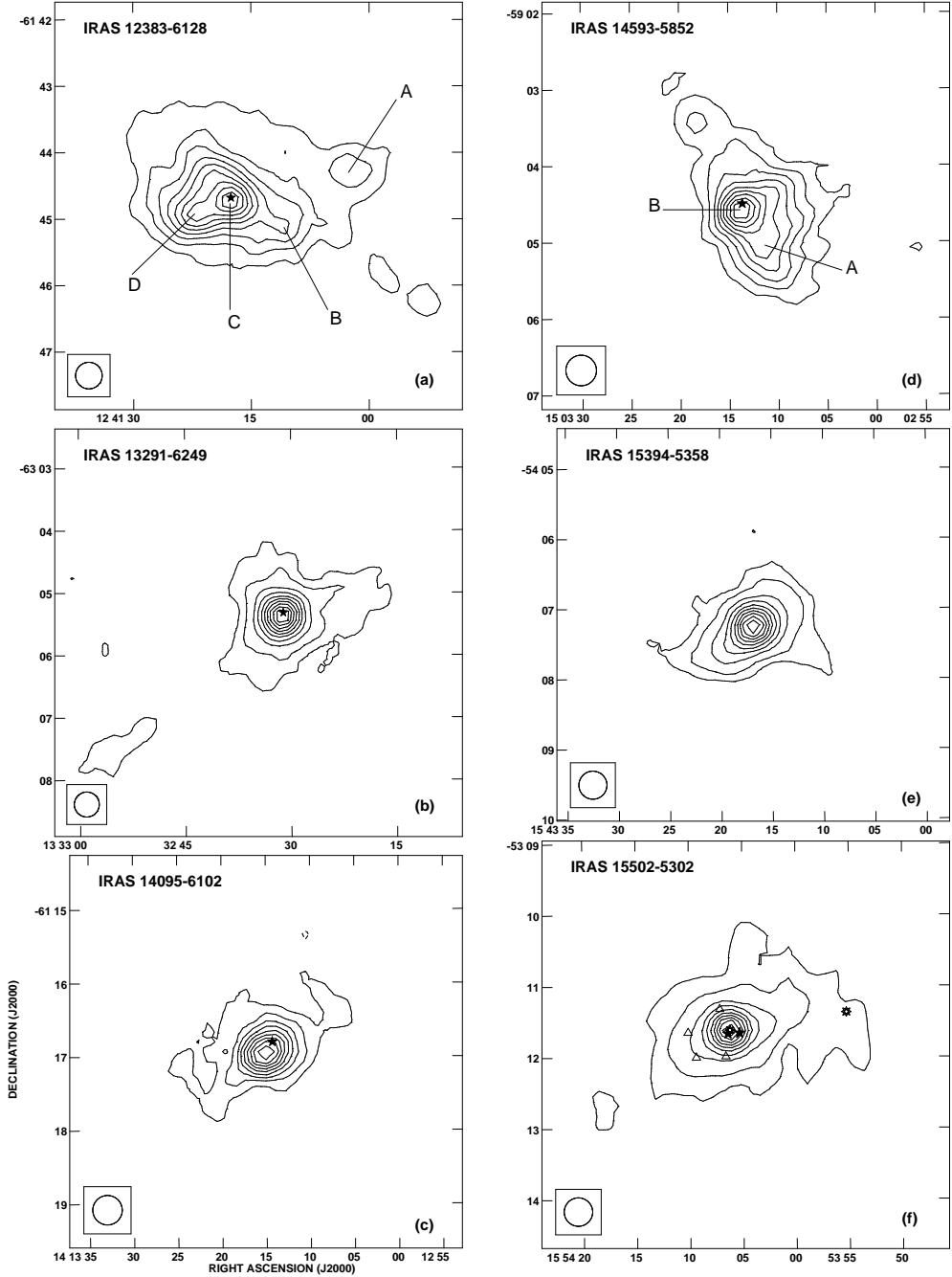}
\vspace{-30.0mm}
\caption
{\baselineskip3.0pt
SEST/SIMBA maps of the 1.2-mm dust continuum emission towards 
young massive star forming regions. FWHM beams are shown in the lower 
left corner. Contour levels are either: (i) -10, 10 to 90 by 10 percent; 
(ii) -5, 5, 10 to 90 by 10 percent; or (iii)
-2.5, 2.5, 5, 10 to 90 by 10 percent, of the observed peak flux density. 
(a) IRAS 12383$-$6128. Contour levels (i). 
Peak flux density $1.15$ Jy beam$^{-1}$; 
(b) IRAS 13291-6249. Contour levels (ii). 
Peak flux density $2.33$ Jy beam$^{-1}$; 
(c) IRAS 14095-6102. Contour levels (i).
Peak flux density $1.99$ Jy beam$^{-1}$;
(d) IRAS 14593-5852. Contour levels (i).
Peak flux density $1.79$ Jy beam$^{-1}$;
(e) IRAS 15394-5358. Contour levels (ii).
Peak flux density $5.18$ Jy beam$^{-1}$;
(f) IRAS 15502-5302. Contour levels (ii).
Peak flux density $5.57$ Jy beam$^{-1}$. 
\label{fig-simba1}}
\end{figure}

{\sl IRAS 12383-6128 (G301.731+1.104)}.$-$The 1.2-mm emission towards IRAS 
12383-6128 (see Fig.~\ref{fig-simba1}a)
arises from an extended region with an irregular morphology, showing at least 
four peaks. The whole region has deconvolved major and minor axis of, respectively, 
116\arcsec\ and 72\arcsec, and a total flux density of 13.4 Jy. 
The morphology of the region was best fit as being produced by four compact 
components (labeled A, B, C and D) and an extended halo. 
The sizes and radius of the individual components are given in 
Table~\ref{tbl-obsparind}. The compact cometary-like HII region detected with ATCA 
lies at the center of the dust core and is associated with the brighter
1.2-mm component (C).
 
{\sl IRAS 13291-6249 (G307.560-0.586)}.$-$The 1.2-mm emission towards IRAS 
13291-6249 (see Fig.~\ref{fig-simba1}b) arises from a bright compact component 
and a weak, extended, irregular component. The total flux density of the 
region is 8.5 Jy. The compact cometary-like HII region detected with ATCA lies 
at the peak position of the dust core.

{\sl IRAS 14095-6102 (G312.596+0.048)}.$-$The 1.2-mm emission towards IRAS 
14095-6102 (see Fig.~\ref{fig-simba1}c) arises from a single structure, with 
a {\it c+e} morphology. The total flux density of the region is 6.7 Jy. The 
compact HII region detected with ATCA lies near the peak position of the dust core.

{\sl IRAS 14593-5852 (G319.163-0.419)}.$-$The 1.2-mm emission towards IRAS 
14593-5852 (see Fig.~\ref{fig-simba1}d) arises from an unresolved bright compact 
component (labeled B), with a flux 
density of 0.88 Jy, and a weaker extended component (labeled A) with deconvolved 
major and minor axis of 90\arcsec\ and 54\arcsec, respectively. The total 
flux density of the region is 11.3 Jy. The compact HII region detected with 
ATCA lies at the peak position of the dust core.

{\sl IRAS 15394-5358 (G326.474+0.697)}.$-$The 1.2-mm emission towards IRAS 
15394-5358 (see Fig.~\ref{fig-simba1}e) arises from a single structure, with 
a {\it c+e} morphology. The total flux density of the region is 15.9 Jy. No 
radio continuum emission was detected with ATCA towards this dust core up 
to a $3\sigma$ level of 0.5 mJy at 8.6 GHz.

{\sl IRAS 15502-5302 (G328.307+0.432)}.$-$The 1.2-mm emission towards IRAS 
15502-5302 (see Fig.~\ref{fig-simba1}f) arises from a bright compact component 
and a weaker extended component. The later has deconvolved major and minor axis 
of, respectively, 131\arcsec\ and 73\arcsec. The total flux density of the 
region is 27.3 Jy.  The two compact HII regions detected with ATCA (radio 
components A and B; peak indicated by filled stars) lie near the peak position 
of the compact component, which is also associated with the brighter of the two 
MSX-E sources detected towards the region. The four more diffuse HII regions 
(radio components C, D, E, and F; peak positions indicated by open triangles) 
are located within the extended component. The weak protuberance of 1.2-mm 
emission seen to the west is associated with the second MSX-E source.

\begin{figure}
\epsscale{1.0}
\plotone{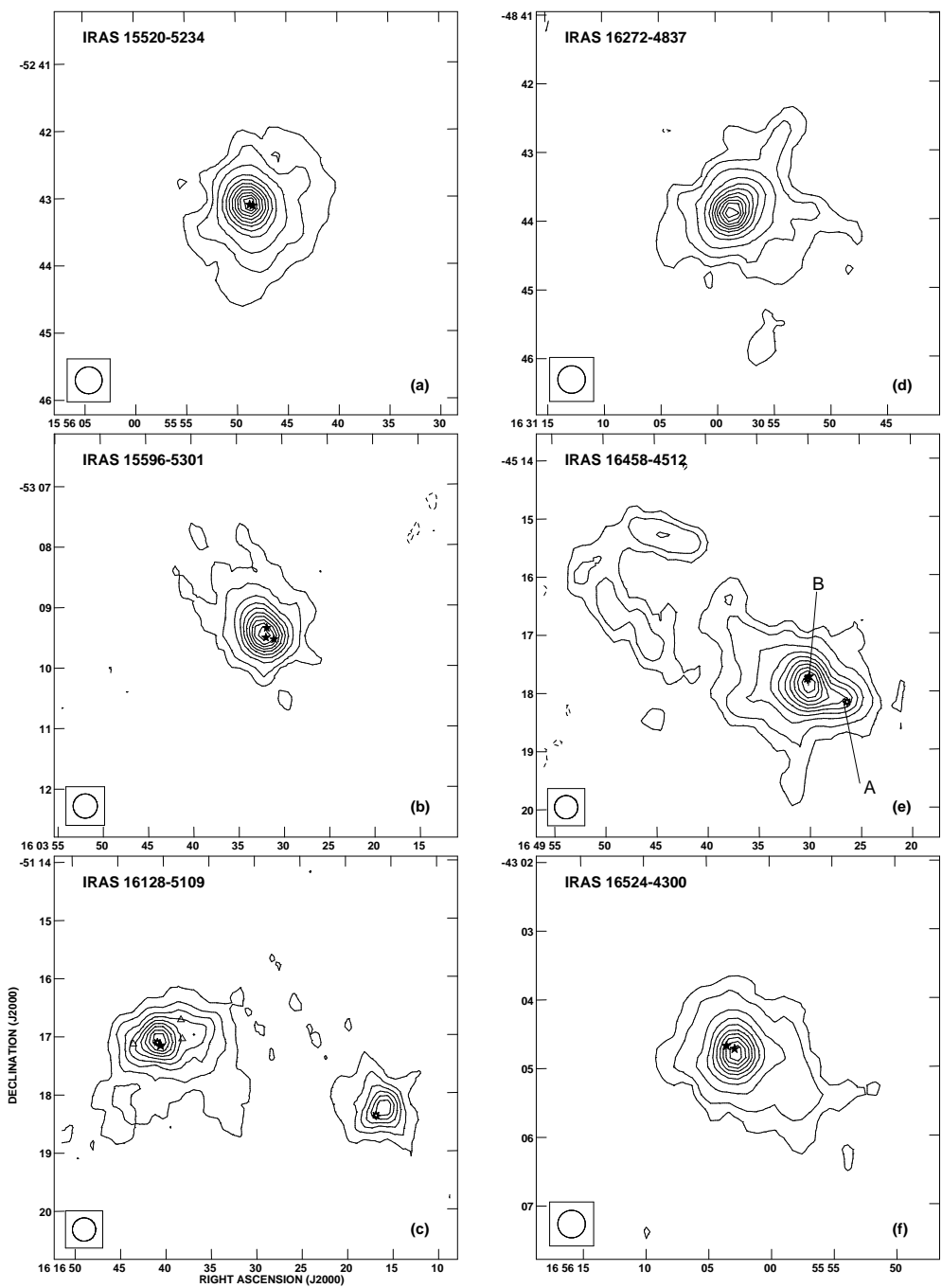}
\vspace{-30.0mm}
\caption
{\baselineskip3.0pt
Same as in Fig. 1.
Panel (a) IRAS 15520-5234. Contour levels (iii).
Peak flux density $9.92$ Jy beam$^{-1}$.
Panel (b) IRAS 15596-5301. Contour levels (ii).
Peak flux density $2.45$ Jy beam$^{-1}$.
Panel (c) IRAS 16128-5109. Contour levels (i).
Peak flux density $2.74$ Jy beam$^{-1}$.
Panel (d) IRAS 16272-4837. Contour levels (iii).
Peak flux density $5.09$ Jy beam$^{-1}$. 
Panel (e) IRAS 16458-4512. Contour levels (ii).
Peak flux density $2.31$ Jy beam$^{-1}$. 
Panel (f) IRAS 16524-4300. Contour levels (ii).
Peak flux density $2.46$ Jy beam$^{-1}$. 
\label{fig-simba2}}
\end{figure}

{\sl IRAS 15520-5234 (G328.808+0.632)}.$-$The 1.2-mm emission towards IRAS 
15520-5234 (see Fig.~\ref{fig-simba2}a) arises from a single structure 
exhibiting a {\it c+e} morphology. The extended emission has deconvolved major 
and minor axis of 100\arcsec\ and 77\arcsec, respectively. The total flux 
density of the region is 27.0 Jy. The two compact HII regions detected with 
ATCA lie at the peak position of the dust core.

{\sl IRAS 15596-5301 (G329.404-0.457)}.$-$The 1.2-mm emission towards IRAS 
15596-530 (see Fig.~\ref{fig-simba2}b) arises from a single structure, with 
roughly a {\it c+e} morphology. The total flux density of the region is 8.3 Jy. 
The three compact HII regions detected with ATCA (Garay et al. 2002) lie near 
the peak position of the dust core.

{\sl IRAS 16128-5109 (G332.153-0.445)}.$-$The 1.2-mm emission within the 
region shown in panel (c) of Fig.~\ref{fig-simba2} arises from two distinct 
sources. The eastern source, associated with IRAS 16128-5109, exhibits a 
{\it c+e} morphology. The extended emission has deconvolved major and minor 
axis of 106\arcsec\ and 77\arcsec, respectively. 
The entire eastern source has a total flux density of 19.9 Jy.  
The brighter component of the complex region of ionized gas detected with 
ATCA (peak indicated by a filled star) lies at the peak of the 
core, which is also associated with the brighter of the two MSX-E sources 
detected towards the region (indicated by asterisks). The three more 
diffuse HII regions (radio components a, b, and d; peak positions indicated by 
open triangles) are located within the halo. 
The western 1.2-mm source, referred as 16128-West in Table~\ref{tbl-obsparwhole},
exhibits a Gaussian morphology, with deconvolved major and minor axis of
48\arcsec\ and 41\arcsec, respectively, has a flux density of 6.8 Jy,
and it is associated with the second MSX-E source detected in the field.   
 
{\sl IRAS 16272-4837 (G335.582-0.284)}.$-$The 1.2-mm emission towards IRAS 
16272-4837 (see Fig.~\ref{fig-simba2}d) arises from a single structure 
consisting of a bright central component and a weak, extended irregular 
component showing elongations to the west and north-west. The total flux 
density of the region is 13.9 Jy. No radio continuum emission was detected 
towards this core up to a level of 0.2 mJy at 4.8 GHz (Garay et al. 2002).

{\sl IRAS 16458-4512 (G340.248-0.373)}.$-$The 1.2-mm emission within the 
region shown in panel (e) of Fig.~\ref{fig-simba2} arises from two distinct 
extended regions. The region to the southwest, associated with IRAS 16458-4512, 
exhibits an irregular morphology, with at least two components (labeled A and B).
The fitted parameters of the individual components are given in
Table~\ref{tbl-obsparind}. Both components are associated with energy sources 
detected by MSX in the E band. The compact HII region detected with ATCA lies 
at the peak position of the brighter 1.2-mm component (component B), indicating 
it harbors a more luminous object than component A. 

The northeast region, which exhibits a C-shape like morphology, 
is not associated with MSX sources and appears in silhouette against the 
background emission in the MSX-A band, suggesting it corresponds to a 
cold filamentary structure with large column densities.

{\sl IRAS 16524-4300 (G342.704+0.130)}.$-$The 1.2-mm emission towards IRAS 
16524-4300 (see Fig.~\ref{fig-simba2}f) arises from a single structure with 
a {\it c+e} morphology. The extended emission has deconvolved major and minor 
axis of, respectively, 122\arcsec\ and 80\arcsec. The total flux density of 
the region is 11.1 Jy. The two compact HII regions detected with ATCA lie 
near the peak position of the bright compact component.

\begin{figure}
\epsscale{1.0}
\plotone{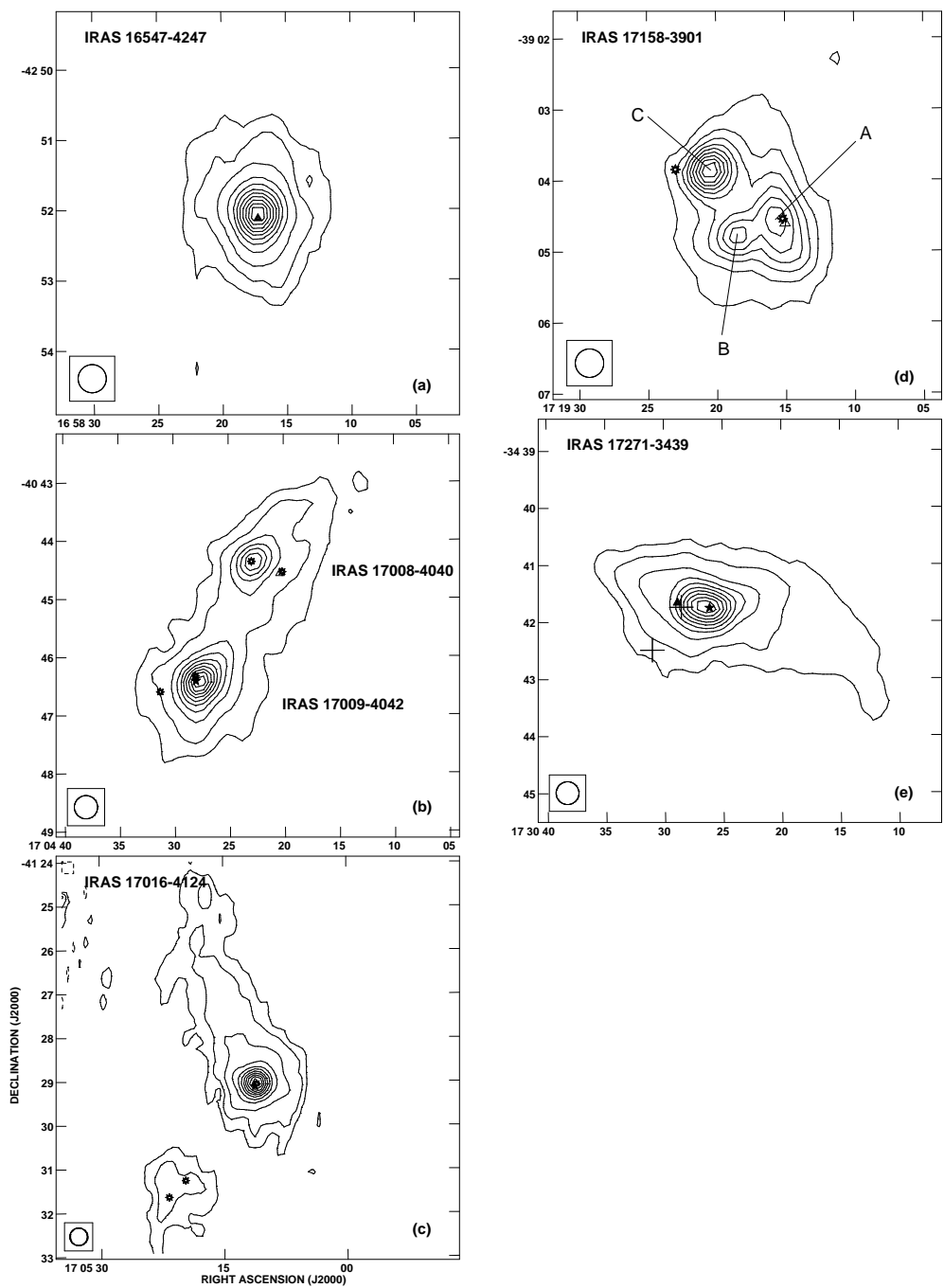}
\vspace{-30.0mm}
\caption
{\baselineskip3.0pt
Same as in Fig. 1.
Panel (a) IRAS 16547-4247. Contour levels (iii). 
Peak flux density $7.33$ Jy beam$^{-1}$.
Panel (b) IRAS 17008-4040 and IRAS 17009-4042. Contour levels (ii).
Peak flux density $9.38$ Jy beam$^{-1}$.
Panel (c) IRAS17016-4124. Contour levels (iii).
Peak flux density $8.42$ Jy beam$^{-1}$.
Panel (d) IRAS 17158-3901. Contour levels (i).
Peak flux density $2.97$ Jy beam$^{-1}$.
Panel (e) IRAS 17271-3439. Contour levels (ii).
Peak flux density $9.86$ Jy beam$^{-1}$.
\label{fig-simba3}}
\end{figure}

{\sl IRAS 16547-4247 (G343.126-0.062)}.$-$The 1.2-mm emission towards IRAS 
16547-4247 (see Fig.~\ref{fig-simba3}a) arises from a single structure which 
exhibits a {\it c+e} morphology. The total flux density of the region is 
20.8 Jy. Radio continuum observations towards this core show the presence of 
a triple radio source. The central radio component (shown as a triangle) 
correspond to a thermal jet and is located at the peak position of the dust core. 

{\sl IRAS 17008-4040 and 17009-4042 (G345.499+0.354 and G345.490+0.311)}.$-$The 
1.2-mm emission within the region shown in panel (b) of Fig.~\ref{fig-simba3}
arises from an extended elongated structure exhibiting two components. 
The northern component, associated with IRAS 17008-4040, shows a {\it c+e} 
morphology. It has deconvolved major and minor axis of, respectively,
102\arcsec\ and 57\arcsec, and a flux density of 34.1 Jy. 
The brighter of the two MSX-E sources detected 
near the northern component lies at the peak position of the 1.2-mm emission.
No radio emission was detected towards this peak position up 
to a 3$\sigma$ level of 1.5 mJy at 4.8 GHz. The extended HII region detected 
with ATCA towards IRAS 17008-4040 (peak position indicated by an open triangle) 
is displaced towards the east from the peak position of the 1.2-mm 
emission, by $\sim30\arcsec$, and is coincident with the second MSX-E source.
 
The southern 1.2-mm component, associated with IRAS 17009-4042, 
exhibits a Gaussian brightness distribution with deconvolved 
major and minor axis of 49\arcsec\ and 34\arcsec, respectively, and a flux 
density of 44.8 Jy. The compact HII region detected with ATCA lies at the peak 
position of the 1.2-mm emission, which is also associated with the brighter of 
the two MSX-E sources detected near this component.

{\sl IRAS 17016-4124 (G345.001-0.220)}.$-$The 1.2-mm emission within the 
region shown in panel (c) of Fig.~\ref{fig-simba3} arises from two distinct 
sources. The northern object, with a total flux density of 45.0 Jy, consists 
of an extended filamentary structure with a bright compact component at its 
southern end. The filament has deconvolved major and minor axis of 308\arcsec\ 
and 95\arcsec, respectively, and a flux density of 33.9 Jy. The compact core,
associated with IRAS 17016-4124, has deconvolved major and minor axis of 
24\arcsec\ and 21\arcsec, respectively, and a flux density of 13.7 Jy. 
The compact HII region detected with ATCA lies at the peak position of the
compact core, which is also associated with the brighter of the three MSX-E 
sources detected within the region.

The southern 1.2-mm object, referred as 17016-South in Table~\ref{tbl-obsparwhole},
exhibits an irregular morphology, with deconvolved major and minor axis of 
89\arcsec\ and 85\arcsec, respectively, and a flux density of 6.5 Jy. It is 
associated with the other two MSX-E objects detected within the region.

{\sl IRAS 17158-3901 (G348.534-0.973)}.$-$The 1.2-mm emission towards IRAS 
17158-3901 (see Fig.~\ref{fig-simba3}d) show complex morphology, which can be 
decomposed as arising from at least three components. The total flux density of 
the entire source is 26.7 Jy. The western component (labeled A) has deconvolved 
major and minor axis of 179\arcsec\ and 35\arcsec, respectively, and a flux 
density of 11.2 Jy. Component A is associated with the HII region detected with 
ATCA and with the brighter of the two MSX-E band sources in the field.
The southern component (labeled B) has deconvolved major and minor
axis of 76\arcsec\ and 49\arcsec, respectively, and a total flux density of
11.5 Jy. The eastern component (labeled C), the brighter of the 1.2-mm 
components (peak flux of 3.0 Jy/beam), has deconvolved major and minor
axis of 30\arcsec\ and 24\arcsec, respectively, and a total flux density of
6.3 Jy. It is associated, although not coincident, with the second MSX-E source 
detected in the field.

{\sl IRAS 17271-3439 (G353.410-0.367)}.$-$The 1.2-mm emission toward IRAS 
17271$-$3439 (see Fig.~\ref{fig-simba3}e) arises from a bright compact central 
component and an elongated weak extended component. The total flux density 
measured within the whole region is 89 Jy. The 
compact component has deconvolved major and minor angular diameters of 37\arcsec\ 
and 21\arcsec, respectively, and a flux density of 18 Jy. The extended 
component has deconvolved major and minor angular diameters of 169\arcsec\ and 
84\arcsec, and a flux density of 65 Jy. The ultra compact 
HII region detected with ATCA is coincident, within the errors, with the 1.2-mm 
peak position. 

Projected toward the dust cloud lie the two 1.4 GHz radio sources (peak 
positions marked by the crosses) detected by the NRAO VLA Sky Survey
(NVSS; Condon et al. 1998) within the $7\times7$ 
\arcmin\ region of the sky shown in Fig.~\ref{fig-simba3}e. The northern NVSS 
source corresponds to the shell-like HII region detected with ATCA at 1.4 GHz, 
and is associated with the IRAS source (peak position marked by the triangle). 
The southern NVSS source is an extended HII region, with angular diameters 
of 88.7\arcsec $\times$74.0\arcsec\, and is resolved out in our ATCA radio 
observations (paper I).

%%%%%%%%%%%%%%%%%%%%%%%%%%%%%%%%%%%%%%%%%%%%%%%%%%%%%%%%%%%%%%%%%%%%%%%%%
% DISCUSSION
%%%%%%%%%%%%%%%%%%%%%%%%%%%%%%%%%%%%%%%%%%%%%%%%%%%%%%%%%%%%%%%%%%%%%%%%%

\section{Discussion}

\subsection{Structure of the dust cores} 

Most of the 1.2-mm sources associated with the IRAS objects exhibit 
single-peaked structures. In only four cases (IRAS 12383-6128, 14593-5852, 
16458-4512, and 17158-3901) we see evidence for the presence of multiple 
components. To 
discern whether the lack of clustering is due to intrinsic reasons, such as the 
absence of fragmentation, or due to the coarse angular resolution of the present 
observations requires observations with high angular resolution. 

\begin{figure}
\epsscale{1.0}
\plotone{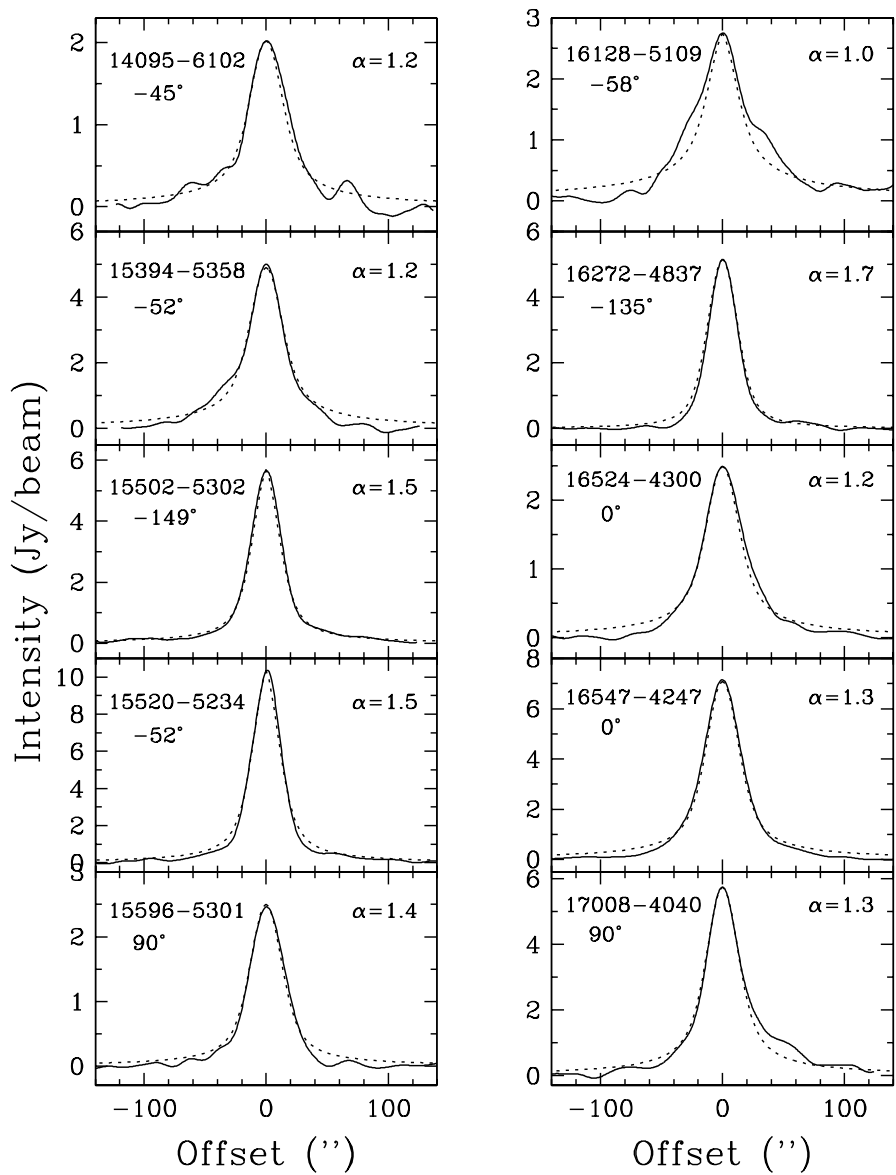}
\vspace{-30.0mm}
\caption
{\baselineskip3.0pt
Intensity cuts across 1.2-mm sources with core-halo morphologies. 
The source name and position angle of the slice, which passes through the 
source peak position, are given in the upper left corner. Dotted lines 
correspond to fits of the observed radial intensity with single power-law 
intensity profiles. The fitted power-law indices are given in the upper right 
corner.
\label{fig-intcuts}}
\end{figure}

An important observational evidence that bears on the physical structure 
of the cores, is the fact that most of the single-peaked objects show 
{\it c+e} morphologies, with clear centrally condensed peaks. This 
is illustrated in Figure~\ref{fig-intcuts} which shows slices of the 
observed 1.2-mm intensity across the ten sources with {\it c+e} 
morphologies. We find that the observed radial 
intensity profiles can be well fitted (see dotted lines) with single power-law 
intensity profiles of the form $I \propto r^{-\alpha}$, where $r$ is the 
distance from the center, properly convolved with the beam of 24\arcsec. The 
fitted power-law indices are in the range from 1.0 to 1.7. 

Assuming that cores have density and temperature radial distributions 
following power laws, then for optically thin dust emission the intensity index 
${\alpha}$ is related to the density index $p$ ($n \propto r^{-p}$) and the 
temperature index $q$ ($T \propto r^{-q}$), by the expression 
$\alpha = p + Qq -1 $ (Adams 1991; Motte \& Andr\'e 2001), where Q is a 
temperature and frequency correction factor with a value of 
$\sim1.2$ at 1.2-mm and 30 K (Beuther et al. 2002). Most of the massive and dense 
cores investigated here are heated by a luminous energy source embedded at their 
central position, as signposted by the presence of either a compact \hii\ region 
or an MSX-E source. van der Tak et al. (2000) have found that for sources
with similar characteristics the temperature decreases with distance 
following a power-law with an index of 0.4. Adopting this value for $q$, we 
then infer, using the above expression, that massive and dense cores have density 
distributions with power-law indices in the range $1.5-2.2$, with an average 
value of 1.8. 

Several dust continuum studies have already shown the presence of steep density 
gradients within massive and dense cores (e.g., van der Tak et al. 2000, 
Mueller et al. 2002, Beuther et al. 2002, Williams et al. 2005). The density 
dependence with radius can be approximated by power-law distributions with 
average values of $p$ in the range between 1.3 and 1.8, similar to the
values derived in this work. The individual values of $p$ exhibit, however, a 
large spread, which is likely to reflect the presence of clumpiness and 
fragmentation within massive and dense cores (e.g., Molinari et al. 2002; 
Beuther \& Schilke 2004). Differences in the fragmentation process of varying 
cores has been already pointed out by Beuther et al. (2005).

\subsection{Overall parameters of the dust cores}

The parameters derived from the 1.2-mm observations are summarized in 
Table~\ref{tbl-derived}. As distances [col.~(2)] we adopted the kinematic 
distances provided by Bronfman (2006, private communication) determined 
using a rotation curve with an orbital velocity of 220 \kms\ at 8.5~kpc from 
the Galactic center (Brand 1986). 
The dust temperatures given in col. (3) 
correspond to the colder temperature determined from fits to the spectral 
energy distribution (SED) which included flux densities at wavelengths 
of 12, 25, 60 and 100 $\mu$m (IRAS data), 8.3, 12.1, 14.7, and 21.3 $\mu$m 
(MSX data), 3.6, 4.5, 5.8 and 8.0 $\mu$m (Spitzer data), and 1.2-mm (SIMBA data).
We find that to achieve a good fit of the SED in this wide 
wavelength range (4 to 1200 $\mu$m) at least three temperature components 
are needed (Morales et al. 2007; paper III).
For the single peaked cores, the radius given 
in col. (4) were computed from the geometric mean of the deconvolved major 
and minor angular sizes obtained from single Gaussian fits to the observed 
spatial structure. The masses [col.~(5)] were computed 
following Chini, Kr\"ugel, \& Wargau (1987), adopting a dust opacity at 1.2-mm 
of 1 cm$^2$ g$^{-1}$ (Ossenkopf \& Henning 1994), a dust-to-gas mass ratio of 
$0.01$, and using the dust temperatures derived from the fit of the SED. 
Cols.~(6) and (7) give, respectively, the average column densities and average  
molecular densities derived from the masses and radius assuming that 
the cores have uniform densities.  Clearly, this is a rough simplification, 
since as discussed in section 4.1 massive and dense cores are likely to have 
steep density gradients.  Finally, the continuum optical depth at 1.2-mm is 
given in col.~(8). 

\begin{figure}
\epsscale{1.0}
\plotone{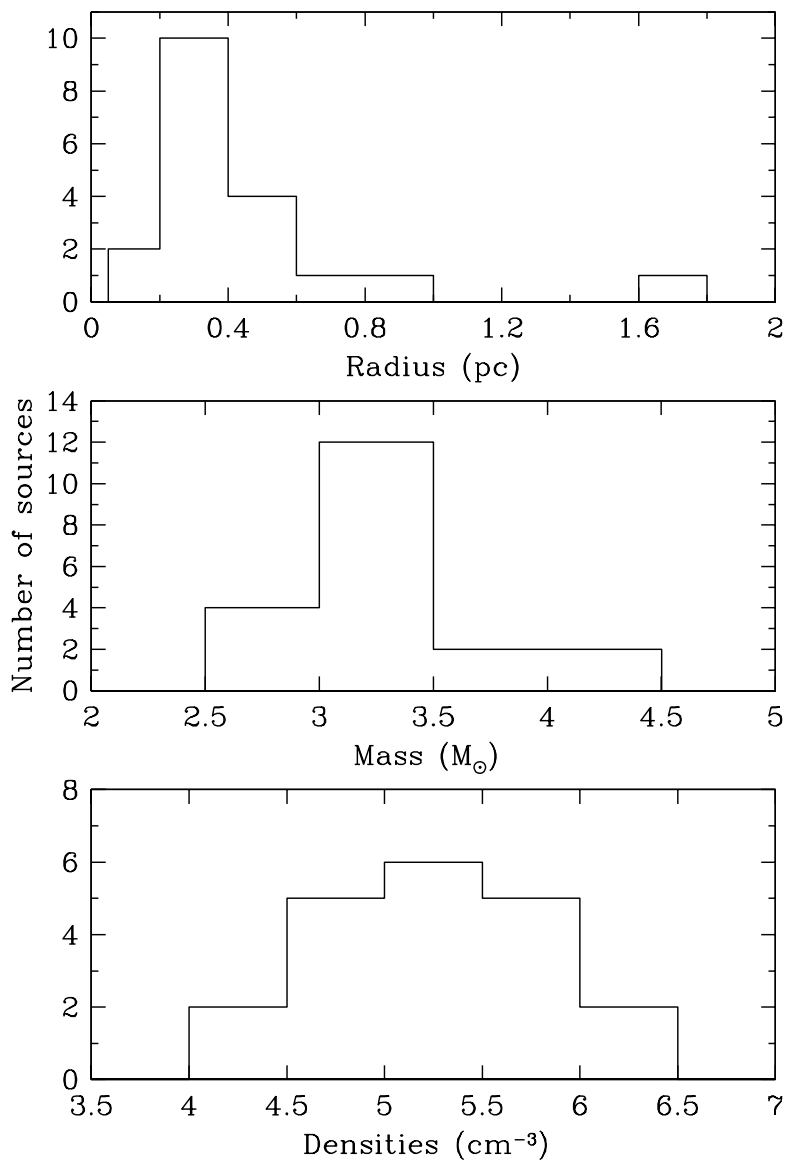}
\vspace{-30.0mm}
\caption
{Distribution of parameters of 1.2-mm cores associated with young high-mass 
star forming regions. Top: Size distribution. Average and median radius are  
0.38 and 0.36 pc, respectively. Middle: Mass distribution. Average and
median mass are $2.4\times10^{3}$ and $1.8\times10^{3}$ M$_{\odot}$, 
respectively. Bottom: Density distribution. Average and median density are 
$3.5\times10^{5}$ and $1.4\times10^{5}$ cm$^{-3}$, respectively.
\label{fig-histopar}}
\end{figure}

Figure~\ref{fig-histopar} shows a histogram with the distribution of the radius, 
masses, and densities of the cores in our sample. We find that the dust cores 
harboring recently formed massive stars have typically sizes of $\sim0.4$ pc, 
masses of $\sim2\times 10^3$ M$_\odot$, column densities of $\sim3\times10^{23}$ 
cm$^{-2}$, densities of $\sim4\times10^5$ cm$^{-3}$, and dust temperatures of 
$30$ K. A comparison with values of other cores containing young massive stars 
derived from molecular line observations (e.g., Cesaroni et al. 1991, Juvela 
1996, Plume et al. 1997) shows good agreement. For instance, from a survey of 
molecular emission, in several transitions of CS, toward massive star forming
regions associated with water masers, Plume et al. (1997) found that high-mass 
stars are formed in molecular cores with typical radii of $0.5$ pc, densities 
of $8\times10^{5}$ cm$^{-3}$, and virial masses of $4\times10^{3}$ M$_{\odot}$. 
Clearly the 1.2-mm dust continuum and high density molecular line emissions are 
tracing the same structures. We conclude that massive stars are formed in 
regions of molecular gas and dust with distinctive physical characteristics, 
which we have referred to as the maternities of massive stars (Garay 2005).

\subsection{High-mass star formation within the massive and dense cores}

Given their large masses, of typically $2\times10^3$ M$_\odot$, the overall 
collapse process of massive and dense cores is likely to produce a protostar 
cluster. The questions of how an individual massive star forms and how the 
cluster forms are therefore closely related. The derived properties of massive 
and dense cores are then important boundary conditions to be taken into account 
for models and simulations. The physical conditions of the massive and dense 
cores with embedded high-mass stars, as the ones investigated here, do in 
fact correspond to the initial conditions for star cluster formation
(Garay et al. 2004), except for the temperatures. In massive and dense cores 
without embedded high-mass stars the temperatures are typically 15 K 
(Menten, Pillai, \& Wyrowski 2005; Shridharan et al. 2005), 
whereas in massive and dense cores with internal energy 
sources the temperatures are typically 32 K (Fa\'undez et al. 2004). 

\begin{figure}
\epsscale{1.0}
\plotone{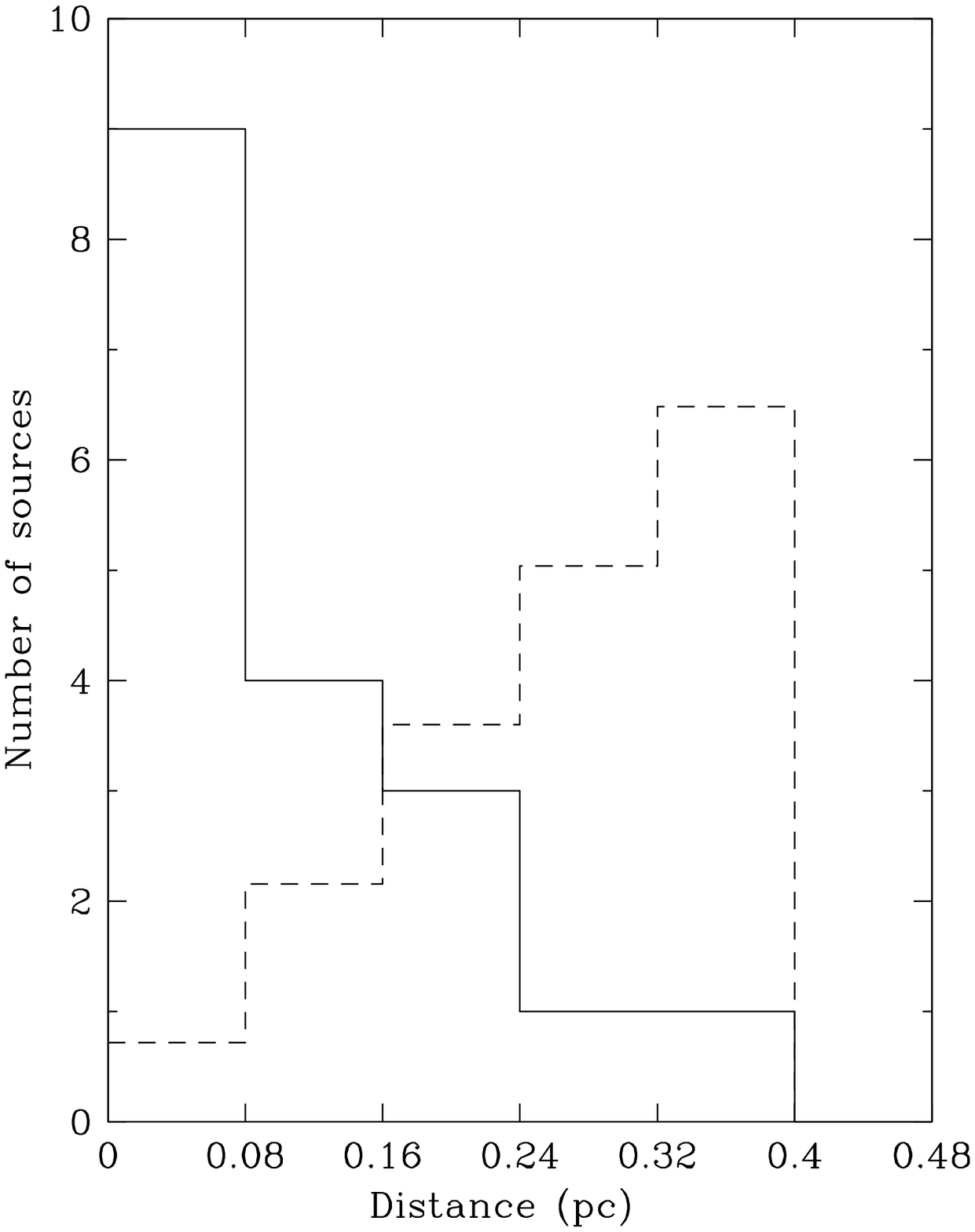}
\vspace{-30.0mm}
\caption
{Distribution of the distance of compact \hii\ regions 
from the peak position of the associated 1.2-mm dust core
(continuos line). The dotted line indicates the expected 
distribution if the regions of ionized gas were uniformly 
distributed across the core. 
\label{fig-histopos}}
\end{figure}

Using the results of the 1.2-mm dust continuum observations reported here and 
the radio continuum observations reported in Paper I, we can address the basic 
question of Where are massive stars formed within protocluster cores? This is 
a relevant question because the location of high-mass stars within massive and 
dense cores may shed some light on their formation mechanism.
Figures~\ref{fig-simba1}-\ref{fig-simba3} strikingly show that in massive 
and dense cores associated with regions of ionized gas, implying that they have 
ongoing massive star formation, the compact \hii\ regions are usually found 
projected at the peak of the dust continuum emission. This is illustrated 
in Figure~\ref{fig-histopos} which shows a histogram of the distribution of 
the linear offsets of the compact \hii\ region from the peak position of the 
associated 1.2-mm core. We find that of the eighteen compact \hii\ regions 
plotted in Figs~\ref{fig-simba1}-\ref{fig-simba3}, twelve are coincident, 
within the errors, with the peak of the 1.2-mm emission and that all them are 
located within 0.4 pc, which is the average radius of the dust cores. 
In addition, the thermal jet associated with IRAS 16547-4247 is also located at 
the peak of the 1.2-mm emission. The dotted line in Fig.~\ref{fig-histopos}
indicates the expected distribution if the compact \hii\ regions were 
uniformly distributed across the dust core. The observed distribution clearly 
shows that massive stars are preferentially born at the center of massive and 
dense cores. 

Two different mechanisms have been proposed to explain the formation of massive
stars: accretion (Osorio, Lizano, \& D'Alessio 1999; Yorke \& Sonnhalter 2002; 
McKee \& Tan 2003) and mergers (Bonnell, Bate, \& Zinnecker 1998; Bonell 2002;
Zinnecker \& Bate 2002). In these two main theoretical ideas, the birth of 
massive stars is envisioned as an event associated with a very dense environment. 
In the merger model the determining parameter is the stellar density whereas 
in the accretion model it is the gas density. In the merger scenario (see review 
by Stahler, Palla, \& Ho 2000) it is proposed that high-mass stars form by the 
merging of low and intermediate mass protostars in a dense cluster environment. 
Observable consequences of both scenarios for massive star formation 
have been thoroughly discussed by Bally \& Zinnecker (2005).

Are the observational data presented in this paper able to discriminate between
the two competing scenarios of high-mass star formation?
In recollection, our observations show that high-mass stars tend to form at the 
center of massive ($\sim2\times10^3$ \Msun), dense ($\sim4\times10^5$ cm$^{-3}$) 
and turbulent ($\Delta v \sim4$ \kms) molecular cores which exhibit steep 
density profiles (typically $n \propto r^{-1.8}$). In addition, 
we find that the virial mass of the cores, determined from molecular line 
observations, and the mass determined from dust continuum observations are in 
very good agreement. This indicates that most of the core mass appear to be in 
the form of molecular gas. 

Lets first consider the merger hypothesis.  
The process of building up massive stars by collisions with lower mass
objects is more efficient at the center of the cluster, where considerable more
dynamical interactions take place. The merger hypothesis then predicts that 
high-mass stars should be found at the center of the cluster, in accord with 
the finding that UC \hii\ regions are usually located at the center of the cores. 
However, for stellar mergers to be responsible for 
the formation of massive stars stellar densities of $\ge10^8$ stars pc$^{-3}$ are 
required (Bonnell 2002). These are more than four orders of magnitude larger 
than the stellar volume densities derived in young embedded dense clusters 
of $\le 10^4$ stars/pc$^3$ (Megeath et al. 1996; Carpenter et al. 1997).
From numerical simulations of the collapse of a $10^3$ \Msun\ fragmenting 
turbulent molecular core, Bonnell at al. (2003) argued that stellar densities of 
$10^6-10^8$ stars pc$^{-3}$ are achieved during a brief period of time 
and suggested that it is during this transient ultradense and highly embedded 
phase that mergers take place. The final state of their simulation (after 
2.4 initial free-fall times) shows a centrally condensed cluster, with the 
fraction of mass in stars being 84\%. This prediction is at 
variance with our observations which shows that even though high-mass stars 
have already been formed at the center of massive cores, most of the 
total mass is still in the form of molecular gas. It is possible that in cores 
with steep density profiles, as the ones investigated here, the fragmentation 
process is not very efficient.
In addition mergers are expected to generate luminous infrared flares, poorly 
collimated impulsive outflows, transient thermal and non-thermal ultracompact 
radio sources (Bally \& Zinnecker 2005), which find little observational
support.
 
The accretion hypothesis requires that the parental cores be dense enough such
that upon collapse the ram pressure of the associated accretion inflow overcame
the radiative forces on dust. McKee \& Tan (2002; 2003) have proposed that massive
stars are formed in centrally condensed turbulent cores. The high degree of 
concentration prevent fragmentation and the accretion rate onto a growing 
protostar at their centers are high enough to overwhelm its radiation pressure 
producing a massive star. 
We find that the cores investigated here have very high mean pressures
(P/k $\sim5\times10^8$ K cm$^{-3}$), large effective sound speeds 
($a_{eff} \sim1.9$ \kms) and high column densities ($3\times10^{23}$ cm$^{-2}$), 
which are the required conditions for high-mass stars to growth by accretion. 
This observational evidence strongly suggest that the cradles of massive stars 
are formed by direct accretion at the center of centrally condensed massive and 
dense cores as hypothesized by McKee \& Tan (2003). In their turbulent and 
pressurized dense core accretion model, the collapse of a massive and dense core 
is likely to produce the birth of a stellar cluster, with most of the mass going 
into relatively low-mass stars. 
The high-mass stars are formed preferentially at the center of the core, 
where the pressure is the highest, and in short time scales of $\sim10^5$ yrs
(Osorio et al. 1999; McKee \& Tan 2002).

With regard to the small observed size of the UC \hii\ regions,
Keto (2002) has shown that a newly formed massive star within a 
dense molecular core may initially produce a confined region of ionized gas. 
The gravitational attraction of the massive central star can maintain 
a steep density gradient and accretion flow within the ionized gas 
and prevents the H II region from expanding hydrodynamically.
In paper I we suggested that the main mechanism of confinement of the 
UC \hii\ regions in our sample, which are excited by stars with an output 
of UV photons of typically $\leq3\times10^{48}$~ s$^{-1}$, is provided by the 
high density and large turbulent pressure of the surrounding molecular gas.
Under these conditions they reach pressure equilibrium in very short times, 
of $\sim5\times10^3$ yrs, having equilibrium radius of typically $\sim0.03$ pc. 
With the currently available data it is not possible to address the question 
of whether or not the formation of high-mass and low-mass stars in the massive 
and dense cores investigated here is coeval. It might be possible that the most 
massive stars are formed first at their center, either as a single object or 
in small groups, followed then by low-mass star formation that is triggered by 
the feedback effects (stellar winds, ionization, compression) of the young 
high-mass stars at the core center. 

%%%%%%%%%%%%%%%%%%%%%%%%%%%%%%%%%%%%%%%%%%%%%%%%%%%%%%%%%%%%%%%%%%%%%%%%%
% CONCLUSIONS
%%%%%%%%%%%%%%%%%%%%%%%%%%%%%%%%%%%%%%%%%%%%%%%%%%%%%%%%%%%%%%%%%%%%%%%%%

\section{Summary}

We made 1.2-mm continuum observations, using SIMBA at the SEST, towards 
eighteen luminous IRAS point sources with colors of UC \hii\ regions and 
CS(2-1) emission. These are thought to be massive star forming regions in 
early stages of evolution. The objectives were to determine the characteristics 
and physical properties of the dust clouds in which high-mass stars form.
Our main results and conclusions are summarized as follows.

Continuum emission was detected toward all the observed targets.
We find that the 1.2-mm sources associated with the luminous IRAS objects,
which we refer to as massive and dense cores, have distinct physical parameters, 
namely linear radius of $\sim0.4$ pc, dust temperatures of $\sim30$ K, masses 
of $\sim2\times10^3$ M$_\odot$, column densities of $\sim3\times10^{23}$ 
cm$^{-2}$, and densities of $\sim4\times10^{5}$ cm$^{-3}$.

Most of the 1.2-mm sources in our sample have single-peaked structures.
In addition, several of them exhibit core-halo morphologies, indicating the 
presence of steep gradients in their densities. 
The observed radial intensity profiles are well modeled with intensity profiles 
following power-law dependences with radius with indices in the range $1.0-1.7$.
This in turn implies that massive and dense cores are centrally condensed, 
with the density decreasing outwards with radius, $r$, as $n\propto r^{-p}$,
with $p$ in the range $1.5-2.2$.

We find that most of the compact \hii\ regions detected with ATCA towards 
our sample of IRAS objects are located at the peak position of the massive and 
dense cores. This is consistent with the hypothesis that the formation of massive 
stars proceeds via accretion in the central regions of very dense and massive 
cores. Under these conditions the regions of ionized gas reach pressure 
equilibrium with their dense and turbulent molecular surroundings in only 
$\sim3\times10^3$ yrs. The high density and large turbulent pressure of the 
molecular gas surrounding UC \hii\ regions is therefore the main mechanism of 
their confinement. 

\acknowledgements

G.G. and D.M. gratefully acknowledge support from the Chilean {\sl Centro de 
Astrof\'\i sica} FONDAP No. 15010003. 

\vfill\eject

%%%%%%%%%%%%%%%%%%%%%%%%%%%%%%%%%%%%%%%%%%%%%%%%%%%%%%%%%%%%%%%%%%%%%%%%%
% REFERENCES
%%%%%%%%%%%%%%%%%%%%%%%%%%%%%%%%%%%%%%%%%%%%%%%%%%%%%%%%%%%%%%%%%%%%%%%%%

\newcommand\rmaap   {RMA\&A~}
\newcommand\aujph   {Aust.~Jour.~Phys.~}
\newcommand\rmaacs   {RMA\&A Conf. Ser.~}

\clearpage

%%%%%%%%%%%%%%%%%%%%%%%%%%%%%%%%%%%%%%%%%%%%%%%%%%%%%%%%%%%%%%%%%%%%%%%%%
% TABLES
%%%%%%%%%%%%%%%%%%%%%%%%%%%%%%%%%%%%%%%%%%%%%%%%%%%%%%%%%%%%%%%%%%%%%%%%%
\newpage
\clearpage

\begin{deluxetable}{llllc}
\tabletypesize{\small}
\tablecolumns{5}
\tablewidth{0pt}
\tablecaption{OBSERVED SOURCES \label{tbl-obssources}}
\tablehead{
\colhead{IRAS source}  & \colhead{Galactic name} & 
\multicolumn{2}{c}{Observed central position}  & \multicolumn{1}{c}{Noise} \\
\cline{3-4}
\colhead{}   &   \colhead{}     & \colhead{$\alpha$(J2000)}      &
\colhead{$\delta$(J2000)}   & \colhead{(mJy beam$^{-1}$)} \\
\colhead{} & \colhead{}   & \colhead{h~~ m~~ s} 
& \colhead{\arcdeg~~ \arcmin~~ \arcsec} & \colhead{} \\
}
\startdata
12383-6128  & G$301.731+1.104$ & 12 41 17.4  & -61 44 40   & 28 \\ 
13291-6249  & G$307.560-0.586$ & 13 32 30.4  & -63 05 19   & 38 \\ 
14095-6102  & G$312.596+0.048$ & 14 13 13.9  & -61 16 48   & 66 \\ 
14593-5852  & G$319.163-0.419$ & 15 03 13.2  & -59 04 24   & 63 \\ 
15394-5358  & G$326.474+0.697$ & 15 43 17.7  & -54 07 30   & 61 \\ 
15502-5302  & G$328.307+0.432$ & 15 54 06.0  & -53 11 38   & 50 \\ 
15520-5234  & G$328.808+0.632$ & 15 55 48.4  & -52 43 10   & 42 \\ 
15596-5301  & G$329.404-0.457$ & 16 03 31.2  & -53 09 29   & 40 \\ 
16128-5109  & G$332.153-0.445$ & 16 16 39.3  & -51 16 58   & 85 \\ 
16272-4837  & G$335.582-0.284$ & 16 30 56.5  & -48 43 46   & 37 \\ 
16458-4512  & G$340.248-0.373$ & 16 49 30.2  & -45 17 50   & 36 \\
16524-4300  & G$342.704+0.130$ & 16 56 04.0  & -43 04 43   & 37 \\
16547-4247  & G$343.126-0.062$ & 16 58 16.9  & -42 52 07   & 45 \\
17008-4040  & G$345.499+0.354$ & 17 04 23.1  & -40 44 26   & 59 \\ 
17009-4042  & G$345.490+0.311$ & 17 04 26.9  & -40 46 27   & 59 \\ 
17016-4124  & G$345.001-0.220$ & 17 05 09.8  & -41 29 04   & 44 \\
17158-3901  & G$348.534-0.973$ & 17 19 16.1  & -39 04 26   & 38 \\
17271-3439  & G$353.416-0.367$ & 17 30 26.5  & -34 41 39   & 40 \\ 
\enddata
\end{deluxetable}

\begin{deluxetable}{lccccccc}
\tablewidth{0pt}
\tablecolumns{8}
\tablecaption{OBSERVED PARAMETERS OF OVERALL 1.2-MM EMISSION 
  \label{tbl-obsparwhole}}
\tablehead{
\colhead{SIMBA}  & \multicolumn{2}{c}{Peak position} & \colhead{} &
 \multicolumn{2}{c}{Flux density\tablenotemark{a}} & \colhead{Angular size
\tablenotemark{b}} & \colhead{Morphology}\\
\cline{2-3} \cline{5-6} 
\colhead{source}                  & \colhead{$\alpha$(2000)}      &
\colhead{$\delta$(2000)}  &  & \colhead{Peak}  & \colhead{Total} & 
\colhead{(\arcsec)} & \colhead{} \\ 
\colhead{}                  & \colhead{}      &
\colhead{} &   & \colhead{(Jy/beam)}  & \colhead{(Jy)} & 
\colhead{} & \colhead{} \\ 
}
\startdata
12383-6128 & 12 41 17.6 & -61 44 43 & & 1.15 & 13.4 & $116\times72$ & multiple \\ 
13291-6249 & 13 32 31.2 & -63 05 23 & & 2.33 &  8.5 & $32\times31$ & compact+irreg.\\
14095-6102 & 14 13 15.0 & -61 16 57 & & 1.99 &  6.7 & $40\times26$  & c+e \\  
14593-5852 & 15 03 13.8 & -59 04 36 & & 1.79 & 11.3 & $81\times50$  & multiple \\  
15394-5358 & 15 43 16.9 & -54 07 15 & & 5.18 & 15.9 & $42\times24$  & c+e \\
15502-5302 & 15 54 06.2 & -53 11 37 & & 5.57 & 27.3 & $48\times31$  & c+e \\
15520-5234 & 15 55 48.8 & -52 43 06 & & 9.92 & 27.0 & $28\times22$  & c+e \\  
15596-5301 & 16 03 32.1 & -53 09 28 & & 2.45 &  8.3 & $39\times28$  & c+e \\
16128-5109 & 16 16 40.5 & -51 17 04 & & 2.74 & 19.9 & $82\times56$  & c+e \\ 
16128-West & 16 16 16.0 & -51 18 16 & & 1.98 &  6.8 & $48\times41$  & Gaussian \\ 
16272-4837 & 16 30 58.7 & -48 43 53 & & 5.09 & 13.9 & $33\times22$  & c+e \\
16458-4512 & 16 49 30.1 & -45 17 52 & & 2.31 & 14.7 & $76\times47$  & multiple \\
16524-4300 & 16 56 02.7 & -43 04 48 & & 2.46 & 11.1 & $47\times45$  & c+e \\ 
16547-4247 & 16 58 17.2 & -42 52 04 & & 7.33 & 20.8 & $34\times25$  & c+e \\
17008-4040 & 17 04 22.8 & -40 44 21 & & 5.77 & 34.1 & $102\times57$ & c+e \\
17009-4042 & 17 04 27.7 & -40 46 26 & & 9.38 & 44.8 & $49\times34$  & Gaussian \\
17016-4124 & 17 05 11.1 & -41 29 03 & & 8.42 & 45.0 & $--$  & core+filament \\
17016-South& 17 05 22.0 & -41 31 14 & & 0.69 &  6.5 & $89\times85$  & irregular\\
17158-3901 & 17 19 20.5 & -39 03 52 & & 2.97 & 26.7 & $129\times73$ & multiple \\ 
17271-3439 & 17 30 26.5 & -34 41 45 & & 9.86 & 89.0 & $87\times49$  & core-halo \\  
\enddata
\tablenotetext{a}{Errors in the flux density are dominated by the 
uncertainties in the flux calibration, of $\sim$20\%.}
\tablenotetext{b}{Errors in the angular sizes are typically 10\%.} 
\end{deluxetable}

\begin{deluxetable}{lcccccc}
\tablewidth{0pt}
\tablecolumns{7}
\tablecaption{OBSERVED PARAMETERS OF INDIVIDUAL 1.2-MM COMPONENTS 
  \label{tbl-obsparind}}
\tablehead{
\colhead{SIMBA}     & \multicolumn{2}{c}{Peak position}  & &
  \multicolumn{2}{c}{Flux density} &  \colhead{Angular size} \\
\cline{2-3} \cline{5-6}
\colhead{source}                  & \colhead{$\alpha$(2000)}      &
\colhead{$\delta$(2000)}  &  & \colhead{Peak}  & \colhead{Total} &
\colhead{(\arcsec)} \\
\colhead{}                  & \colhead{}      &
\colhead{}  &  & \colhead{(Jy/beam)}  & \colhead{(Jy)} &
\colhead{} \\
}
\startdata
12383-6128~A  & 12 41 02.7  & -61 44 19 & & 0.34  & 1.2     &  $59\times41$ \\
12383-6128~B  & 12 41 10.3  & -61 45 05 & & 0.49  & 0.6     &  $37\times18$ \\
12383-6128~C  & 12 41 17.7  & -61 44 38 & & 1.15  & 2.3     &  $57\times20$ \\
12383-6128~D  & 12 41 22.2  & -61 44 56 & & 0.90  & 2.1     &  $46\times28$ \\
12383-6128~halo & 12 41 17.0 & -61 44 26& & --    & $\sim7$ & $212\times116$ \\
14593-5852-A  & 15 03 11.7  & -59 04 42 & & 1.08  & 10.1    &  $90\times54$ \\
14593-5852-B  & 15 03 13.8  & -59 04 36 & & 1.79  & 0.88   & $--$\tablenotemark{a}\\
16458-4512~A  & 16 49 26.2  & -45 18 09 & & 0.82  & 1.31    &  $23\times14$ \\
16458-4512~B  & 16 49 30.4  & -45 17 53 & & 2.31  & 10.3    &  $52\times44$ \\
17016-4124-core & 17 05 11.1 & -41 29 03& & 8.42  & 13.7    &  $24\times21$ \\
17016-4124-fila & 17 05 12.2 & -41 28 28& & 0.80  & 33.9    & $308\times95$ \\
17158-3901~A  & 17 19 15.7  & -39 04 34 & & 2.02  & 11.2    & $179\times35$ \\
17158-3901~B  & 17 19 18.4  & -39 04 47 & & 1.99  & 11.5    &  $76\times49$ \\
17158-3901~C  & 17 19 20.5  & -39 03 52 & & 2.97  &  6.3    &  $30\times24$ \\
17271-3439~core & 17 30 26.5 & -34 41 45& & 9.86  & 18.2    & $37\times21$ \\
17271-3439~halo & 17 30 26.1 & -34 41 40& & --    & 65.4    & $169\times84$ \\
\enddata
\tablenotetext{a}{unresolved.}
\end{deluxetable}

\begin{deluxetable}{lccccccc}
\tablewidth{0pt}
\tablecolumns{8}
\tablecaption{DERIVED PARAMETERS \label{tbl-derived}}
\tablehead{
\colhead{SIMBA} & \colhead{D} & \colhead{T$_d$} & \colhead{Radius} & 
 \colhead{Mass} &  \colhead{n(H$_2$)}  & \colhead{N(H$_2$)} &
 \colhead{$\tau_{1.2mm}$} \\
 \colhead{source} & \colhead{(kpc)} & \colhead{(K)} & \colhead{(pc)} & \colhead{(M$_{\odot}$)} & 
 \colhead{(cm$^{-3})$} & \colhead{(cm$^{-2}$)}  & \colhead{} \\
}
\startdata
12383-6128 &  4.4 & 24. & 0.93 & $3.4\times10^3$ & $1.7\times10^4$ & $6.7\times10^{22}$ & 0.0026 \\ 
13291-6249 &  2.8 & 29. & 0.21 & $5.2\times10^2$ & $2.2\times10^5$ & $1.9\times10^{23}$ & 0.0075 \\ 
14095-6102 &  5.7 & 28. & 0.44 & $2.2\times10^3$ & $1.0\times10^5$ & $1.9\times10^{23}$ & 0.0075 \\ 
14593-5852 & 11.5 & 27. & 1.73 & $1.5\times10^4$ & $1.2\times10^4$ & $8.6\times10^{22}$ & 0.0034 \\ 
15394-5358 &  2.8 & 25. & 0.22 & $1.5\times10^3$ & $5.9\times10^5$ & $5.3\times10^{23}$ & 0.0209 \\ 
15502-5302 &  5.6 & 41. & 0.52 & $4.6\times10^3$ & $1.4\times10^5$ & $2.9\times10^{23}$ & 0.0114 \\ 
15520-5234 &  2.9 & 42. & 0.17 & $1.5\times10^3$ & $1.2\times10^6$ & $8.6\times10^{23}$ & 0.0341 \\ 
15596-5301 &  4.6 & 28. & 0.37 & $1.8\times10^3$ & $1.4\times10^5$ & $2.2\times10^{23}$ & 0.0087 \\ 
16128-5109 &  3.7 & 33. & 0.58 & $2.1\times10^3$ & $4.3\times10^4$ & $1.0\times10^{23}$ & 0.0041 \\ 
16272-4837 &  3.4 & 25. & 0.22 & $2.1\times10^3$ & $8.0\times10^5$ & $7.2\times10^{23}$ & 0.0286 \\ 
16458-4512 &  3.8 & 26. & 0.52 & $2.1\times10^3$ & $6.1\times10^4$ & $1.3\times10^{23}$ & 0.0052 \\ 
16524-4300 &  3.6 & 27. & 0.54 & $1.5\times10^3$ & $4.0\times10^4$ & $8.8\times10^{22}$ & 0.0035 \\ 
16547-4247 &  2.9 & 31. & 0.20 & $1.9\times10^3$ & $9.3\times10^5$ & $7.7\times10^{23}$ & 0.0305 \\ 
17008-4040 &  2.0 & 30. & 0.36 & $1.2\times10^3$ & $1.1\times10^5$ & $1.6\times10^{23}$ & 0.0065 \\ 
17009-4042 &  2.1 & 32. & 0.21 & $1.6\times10^3$ & $7.2\times10^5$ & $6.3\times10^{23}$ & 0.0247 \\ 
17016-4124 &  2.7 & 30. & 0.22 & $1.8\times10^3$ & $6.7\times10^5$ & $6.1\times10^{23}$ & 0.0240 \\ 
17158-A    &  2.0 & 28. & 0.38 & $5.0\times10^2$ & $3.6\times10^4$ & $5.7\times10^{22}$ & 0.0022 \\ 
17158-B    &  2.0 & 28. & 0.29 & $5.1\times10^2$ & $8.1\times10^4$ & $9.9\times10^{22}$ & 0.0039 \\ 
17158-C    &  2.0 & 28. & 0.13 & $2.8\times10^2$ & $5.3\times10^5$ & $2.8\times10^{23}$ & 0.0111 \\ 
17271-3439 &  4.5 & 35. & 0.72 & $1.4\times10^4$ & $1.6\times10^5$ & $4.8\times10^{23}$ & 0.0188 \\ 
\enddata
\end{deluxetable}

\vfill\eject

%%%%%%%%%%%%%%%%%%%%%%%%%%%%%%%%%%%%%%%%%%%%%%%%%%%%%%%%%%%%%%%%%%%%%%%%%
% FIGURES
%%%%%%%%%%%%%%%%%%%%%%%%%%%%%%%%%%%%%%%%%%%%%%%%%%%%%%%%%%%%%%%%%%%%%%%%%

%%%%%%%%%%%%%%%%%%%%%%%%%%%%%%%%%%%%%%%%%%%%%%%%%%%%%%%%%%%%%%%%%%%%%%%%%
%% Use the figure environment and \plotone or \plottwo to include
%% figures and captions in your electronic submission.
%%%%%%%%%%%%%%%%%%%%%%%%%%%%%%%%%%%%%%%%%%%%%%%%%%%%%%%%%%%%%%%%%%%%%%%%%

\begin{references}

\reference{}

\reference{}Adams, F.C. 1991, \apj, 382, 544 

\reference{}Bally, J., \& Zinnecker, H. 2005, \aj, 129, 2281 

\reference{}Beuther, H., \& Schilke, P. 2004, Science, 303, 1167 

\reference{}Beuther, H., Schilke, P., Menten, K. M.,
 Motte, F., Sridharan, T. K., \& Wyrowski, F. 2002, \apj, 566, 945 

\reference{}Beuther, H., Zhang, Q., Sridharan, T. K., \& Chen, Y. 2005, 
 \apj, 628, 800 

\reference{}Bonnell, I.A. 2002, in ASP Conf. Ser. Vol. 267, The Earliest
 Stages of Massive Star Birth, ed. P.A. Crowther (San Francisco: ASP), 193 

\reference{}Bonnell, I.A., Bate, M.R., \& Zinnecker, H. 1998, \mnras, 298, 93 

\reference{}Brand, J. 1986, Ph.D. thesis, Leiden University 

\reference{}Bronfman, L., Nyman, L.{\AA}., \& May, J. 1996, \aaps, 115, 81 

\reference{}Carpenter, J.M., Meyer, M.R., Dougados, C., Strom, S.E.,
   \& Hillenbrand, L.A. 1997, \aj, 114, 198 

\reference{}Cesaroni, R., Walmsley, C.M., K\"{o}mpe, C., \& Churchwell, E.
  1991, \aap, 252, 278

\reference{}Chini, R., Kr\"ugel, E., \& Wargau, W. 1987, \aap, 181, 378 

\reference{}Condon, J.J., Cotton, W.D., Greisen, E.W., Yin, Q.F., Perley, R.A.,
  Taylor, G.B., \& Broderick, J.J. 1998, \aj, 115, 1693

\reference{}Egan, M.P., Shipman, R.F., Price, S.D., Carey, S.J., Clark, F.O.,
  \& Cohen, M. 1998, \apj, 494, L199  

\reference{}Fa\'undez, S., Bronfman, L., Garay, G., Chini, R., Nyman,
  L.-\AA, \& May, J. 2004, \aap, 426, 97 

\reference{}Garay, G. 2005, in Massive star birth: A crossroads of Astrophysics, 
  Proceedings IAU Symposium No. 227, eds. Cesaroni, R., Felli, M., Churchwell, E., 
  Walmsley, M. (Cambridge: Cambridge University Press), 86 

\reference{}Garay, G., Brooks, K.J., Mardones, D., \& Norris, R.P. 2006,
   submitted to \apj. (Paper I)

\reference{}Garay, G., Brooks, K.J., Mardones, D., Norris, R. P.
  \& Burton, M.G. 2002, \apj, 579, 678 

\reference{}Garay, G., Fa\'undez, S., Mardones, D., Bronfman, L., Chini, R., \&
 Nyman, L.-\AA. 2004, \apj, 610, 313 

\reference{}Juvela, M. 1996, \aaps, 118, 191 

\reference{}Keto, E. 2002, ApJ 580, 980

\reference{}Mardones, D. 1998, Ph.D. thesis, Harvard University 

\reference{}McKee, C.F., \& Tan, J. C. 2002, Nature 416, 59 

\reference{}McKee, C.F., \& Tan, J. C. 2003, \apj, 585, 850 

\reference{}Megeath, S.T., Herter, T., Beichman, C., Gautier, N., Hester, J.J.,
    Rayner, J., \& Shupe, D. 1996, \aap, 307, 775 

\reference{}Menten, K.M., Pillai, T., \& Wyrowski, F. 2005, in Massive star 
  birth: A crossroads of Astrophysics, Proceedings IAU Symposium No. 227, eds. 
  Cesaroni, R., Felli, M., Churchwell, E., Walmsley, M. (Cambridge: Cambridge 
  University Press), 23 

\reference{}Molinari, S., Testi, L., Rodr\'\i guez, L.F., \& Zhang, Q. 2002, 
  \apj, 570, 758 

\reference{}Motte, F., \& Andr\'e, P. 2001, \aap, 365, 440 

\reference{}Mueller, K.E., Shirley, Y.L., Evans II, N.J. \& Jacobson, H.R. 2002, 
  \apjs, 143, 469 

\reference{}Osorio, M., Lizano, S., \& D'Alessio, P. 1999, \apj, 525, 808 

\reference{}Ossenkopf, V., \& Henning, Th. 1994, \aap, 291, 943 

\reference{}Plume, R., Jaffe, D.T., \& Evans II, N.J. 1992, \apjs, 78, 505 

\reference{}Plume, R., Jaffe, D.T., Evans II, N.J., Mart\'\i n-Pintado, J. \& 
  G\'omez-Gonz\'alez, J. 1997, \apj,  476, 730 

\reference{}Price, S.D. 1995, \ssr, 74, 81 

\reference{}Shirley, Y.L., Evans II, N.J., Young, K.E., Knez, C., \& Jaffe, D.T. 
  2003, \apjs, 149, 375 

\reference{}Sridharan, T.K., Beuther, H., Saito, M., Wyrowski, F., \& Schilke, P.
2005, \apj, 634, L57

\reference{}Stahler, S.W., Palla, F., \& Ho, P.T.P. 2000, Protostars and Planets 
  IV, ed. V. Mannings, A.P. Boss, \& S.S. Russell (Tucson: Univ. Arizona), 327

\reference{}van der Tak, F.F.S., van Dishoeck, E.F., Evans II, N.J., \&  Blake, G.A. 
  2000, \apj, 537, 283 

\reference{}Williams, S.J., Fuller, G.A., \& Sridharan, T.K. 2004, \aap, 417, 115 

\reference{}Williams, S.J., Fuller, G.A., \& Sridharan, T.K. 2005, \aap, 434, 257

\reference{}Yorke, H.W., \& Sonnhalter, C. 2002, \apj, 569, 846 

\end{references}
\end{document}